\documentclass[twocolumn]{aastex6}

\shorttitle{Maximum Angular Separation}
\shortauthors{Stephen R. Kane et al.}

\begin{document}

\title{Maximum Angular Separation Epochs for Exoplanet Imaging
  Observations}

\author{
  Stephen R. Kane\altaffilmark{1},
  Tiffany Meshkat\altaffilmark{2},
  Margaret C. Turnbull\altaffilmark{3}
}
\altaffiltext{1}{Department of Earth Sciences, University of
  California, Riverside, CA 92521, USA}
\altaffiltext{2}{IPAC, Caltech, M/C 100-22, 1200 East California
  Boulevard, Pasadena, CA 91125, USA}
\altaffiltext{3}{SETI Institute, Carl Sagan Center for the Study of
  Life in the Universe, Off-Site: 2801 Shefford Drive, Madison, WI
  53719, USA}
\email{skane@ucr.edu}

%%%%%%%%%%%%%%%%%%%%%%%%%%%%%%%%%%%%%%%%%%%%%%%%%%%%%%%%%%%%%%%%%%%%

\begin{abstract}

Direct imaging of exoplanets presents both significant challenges and
significant gains. The advantages primarily lie in receiving emitted
and, with future instruments, reflected photons at phase angles not
accessible by other techniques, enabling the potential for atmospheric
studies and the detection of rotation and surface features. The
challenges are numerous and include coronagraph development and
achieving the necessary contrast ratio. Here, we address the specific
challenge of determining epochs of maximum angular separation for the
star and planet. We compute orbital ephemerides for known transiting
and radial velocity planets, taking Keplerian orbital elements into
account. We provide analytical expressions for angular star--planet
separation as a function of the true anomaly, including the locations
of minimum and maximum. These expressions are used to calculate
uncertainties for maximum angular separation as a function of time for
the known exoplanets, and we provide strategies for improving
ephemerides with application to proposed and planned imaging missions.

\end{abstract}


%%%%%%%%%%%%%%%%%%%%%%%%%%%%%%%%%%%%%%%%%%%%%%%%%%%%%%%%%%%%%%%%%%%%

\section{Introduction}
\label{intro}

Direct imaging of exoplanets provides opportunities for significantly
extending exoplanet science, such as direct atmospheric retrieval
\citep{fen18}, and unlocking intrinsic planet properties, such as
albedo, rotation, and obliquity \citep{cow09,kan17}. The method of
direct imaging also remains one of the most challenging techniques for
studying exoplanets. At the present time, only $\sim$1\% of the known
exoplanets have been discovered using direct imaging, according to
data from the NASA Exoplanet Archive \citep{ake13}. However, many
technology advancements, both instrumental and with software, have
taken place over recent years that allow significantly enhanced
capabilities to extract a planetary signature from the stellar
diffraction pattern. The Gemini Planet Imager (GPI) \citep{mac14} and
the Spectro-Polarimetric High-Contrast Exoplanet Research (SPHERE)
\citep{beu08} instruments have contributed significantly to the
ground-based imaged planets inventory. Examples of other current and
planned ground-based instruments include the Subaru Coronagraphic
Extreme Adaptive Optics (SCExAO) instrument \citep{jov16}, the
Magellan Adaptive Optics (MagAO-X) instrument \citep{mal18}, the Keck
Planet Imager and Characterizer \citep{maw17}, and the Planet
Formation Imager \citep{mon18}. Development continues to progress for
proposed space-based imaging facilities, such as the WFIRST
coronagraph \citep{dou18}, the Habitable Exoplanet imaging mission
(HabEx) \citep{ary17}, and the Large Ultraviolet/Optical/Infrared
Surveyor (LUVOIR) \citep{fra17}. The methodology for classifying
discoveries from such facilities and their expected yields is a key
component for the mission science drivers \citep{kop18}.

The observing strategy for direct imaging efforts requires an
efficient target selection and time management, particularly for
space-based resources. For known indirectly detected exoplanets, the
optimal observing times require sufficient orbital architecture
knowledge to constrain when the planet will have an angular separation
from the host star that places it outside of the inner working angle
\citep{kan13,sch15}. Many of the radial velocity (RV) planets, for
example, have poorly determined orbital ephemerides due to the
uncertainties in the Keplerian orbital solution compacted by the time
since last observation \citep{kan09,jen10}. Observing those host stars
with the after a long time baseline can help reacquire the planet's
orbital phase and dramatically improve the ephemerides.

In this paper, we address the issue of determining the maximum angular
separation between the star and planet for Keplerian orbital
solutions. In Section~\ref{orbit} we discuss the challenge of orbital
ephemerides and calculate the uncertainty in orbital location for 300
known exoplanets projected forward to 2025. In Section~\ref{ang}, we
provide analytical expressions for both the star--planet separation
and the derivative with respect to the true anomaly, which allows the
epoch of maximum angular separation to be
determined. Section~\ref{epoch} combines the work of the previous
sections and provides calculated maximum angular separations, orbital
phases where they occur, and uncertainties on those orbital locations
for 50 known exoplanets. We provide concluding remarks in
Section~\ref{conclusions} and recommendations for observing strategies
designed to improve orbital ephemerides for direct imaging
observations.

%%%%%%%%%%%%%%%%%%%%%%%%%%%%%%%%%%%%%%%%%%%%%%%%%%%%%%%%%%%%%%%%%%%%

\section{Exoplanet Orbital Ephemerides}
\label{orbit}

At the present time, exoplanet discoveries are dominated by those that
utilize the transit method. Here we focus on those planets that have
full Keplerian orbital solutions in order to provide a complete
description of the orbital phase and angular separation. Exoplanets
with Keplerian orbital solutions tend to be those discovered with the
RV technique, for which survey durations have extended the period
sensitivity beyond $\sim$10~years \citep{wit16}. Data regarding
exoplanets and orbital parameters are available from numerous sources,
both in the literature and online \citep{but06,wri11}. For this study,
we utilize the data from the NASA Exoplanet Archive \citep{ake13},
where the data are current as of 2018 August 17.

\begin{figure}
  \includegraphics[angle=270,width=8.2cm]{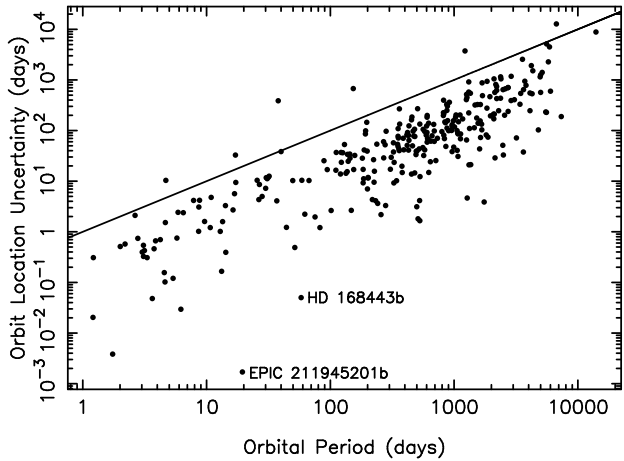}
  \caption{The calculated uncertainty in the planetary orbital
    location as a function of orbital period for 300 known
    exoplanets. The solid line shows where the uncertainty in the
    orbital location is the same as the orbital period. The
    uncertainty in the orbital location has been calculated for
    January 1, 2025 by propagating the uncertainties in the orbital
    period and time of periastron passage.}
  \label{uncfig}
\end{figure}

We calculate the uncertainty in the planetary orbital location for
2025 January 1 (JD = 2,460,676.5). This date was chosen since it
approximately matches the anticipated first light and/or launch of
numerous ground and space-based telescopes that aim to directly image
long-period planets. We propagate the uncertainties from the time of
periastron passage using methodology of \citet{kan09}. This
methodology uses Keplerian orbital elements and their uncertainties
with multiples of the orbital period to calculate epochs of specific
orbit locations relative to times of measured periastron
passage. \citet{kan09} uses this method to determine uncertainties and
transit windows for times of inferior conjunction, whereas we use the
method to describe the general uncertainty on orbital location. The
results of these calculations are shown in Figure~\ref{uncfig}, which
plots the uncertainty in the orbital location as a function of the
orbital period. The relationship between these two parameters follows
a power law where the uncertainty in orbital location becomes
comparable to the orbital period for particularly long periods,
indicated by the solid line. This means that the location of those
planets in their orbit has been completely lost, making it impossible
to provide useful ephemeris information for follow-up observations
that require such knowledge. An example of such follow-up observations
is the need to predict times of maximum angular separation for direct
imaging experiments.

There are several significant outliers in Figure~\ref{uncfig} for
which the ephemerides are relatively well defined. The combination of
RV data with transit data from the {\it K2} mission by \citet{cha18}
produced an exceptionally strong constraint on the time of periastron
passage for EPIC~211945201b. In the case of HD~168443b, targeted RV
observations during periastron by the Transit Ephemeris Refinement and
Monitoring Survey (TERMS), combined with a long time baseline, yielded
very small uncertainties on the time of periastron passage
\citep{pil11}. In general, the calculations presented here demonstrate
the need for further RV observations to reacquire the planetary
location.

%%%%%%%%%%%%%%%%%%%%%%%%%%%%%%%%%%%%%%%%%%%%%%%%%%%%%%%%%%%%%%%%%%%%

\section{Maximum Angular Separation}
\label{ang}

\begin{figure*}
  \begin{center}
    \begin{tabular}{cc}
      \includegraphics[angle=270,width=8.5cm]{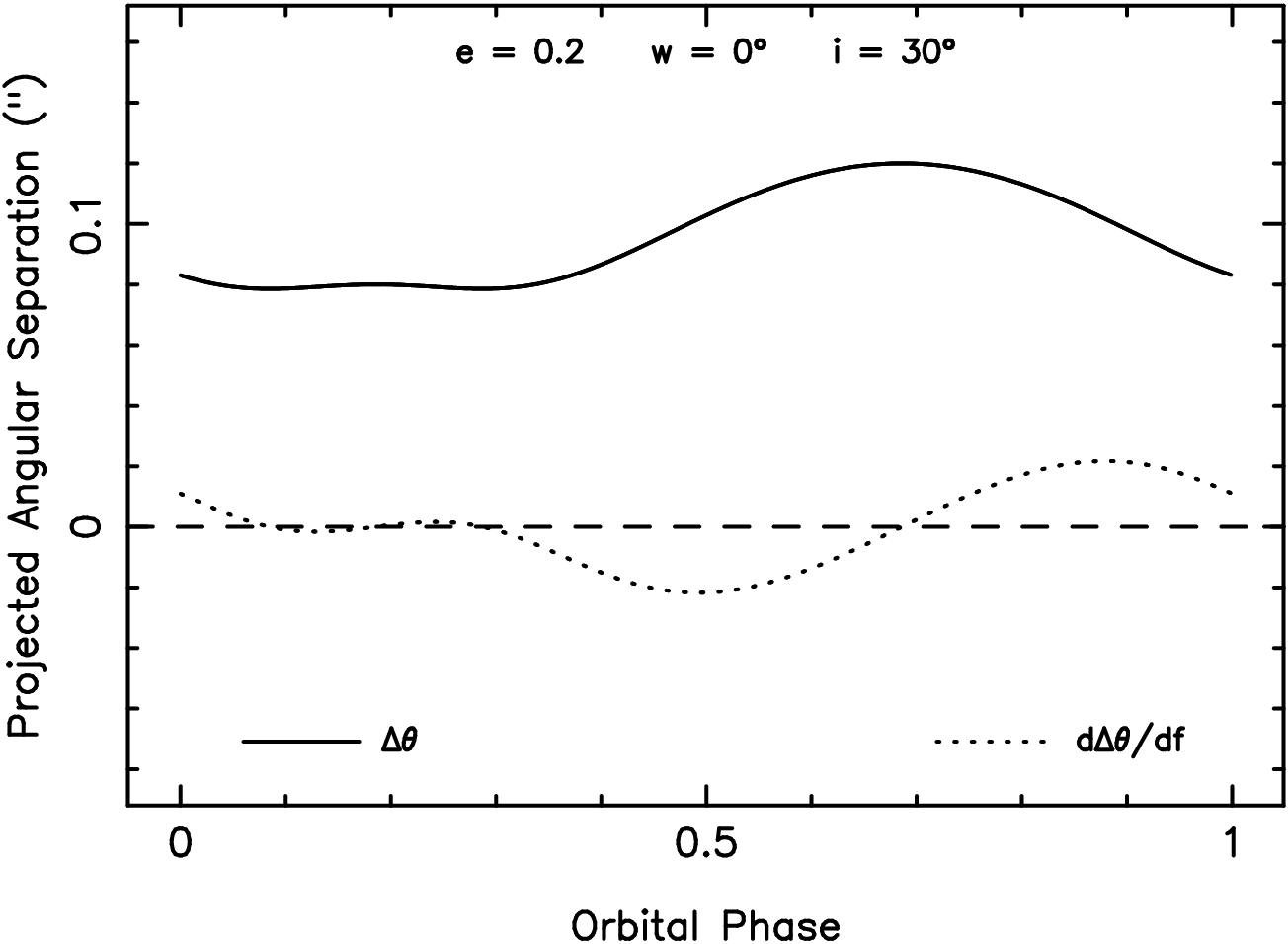} &
      \includegraphics[angle=270,width=8.5cm]{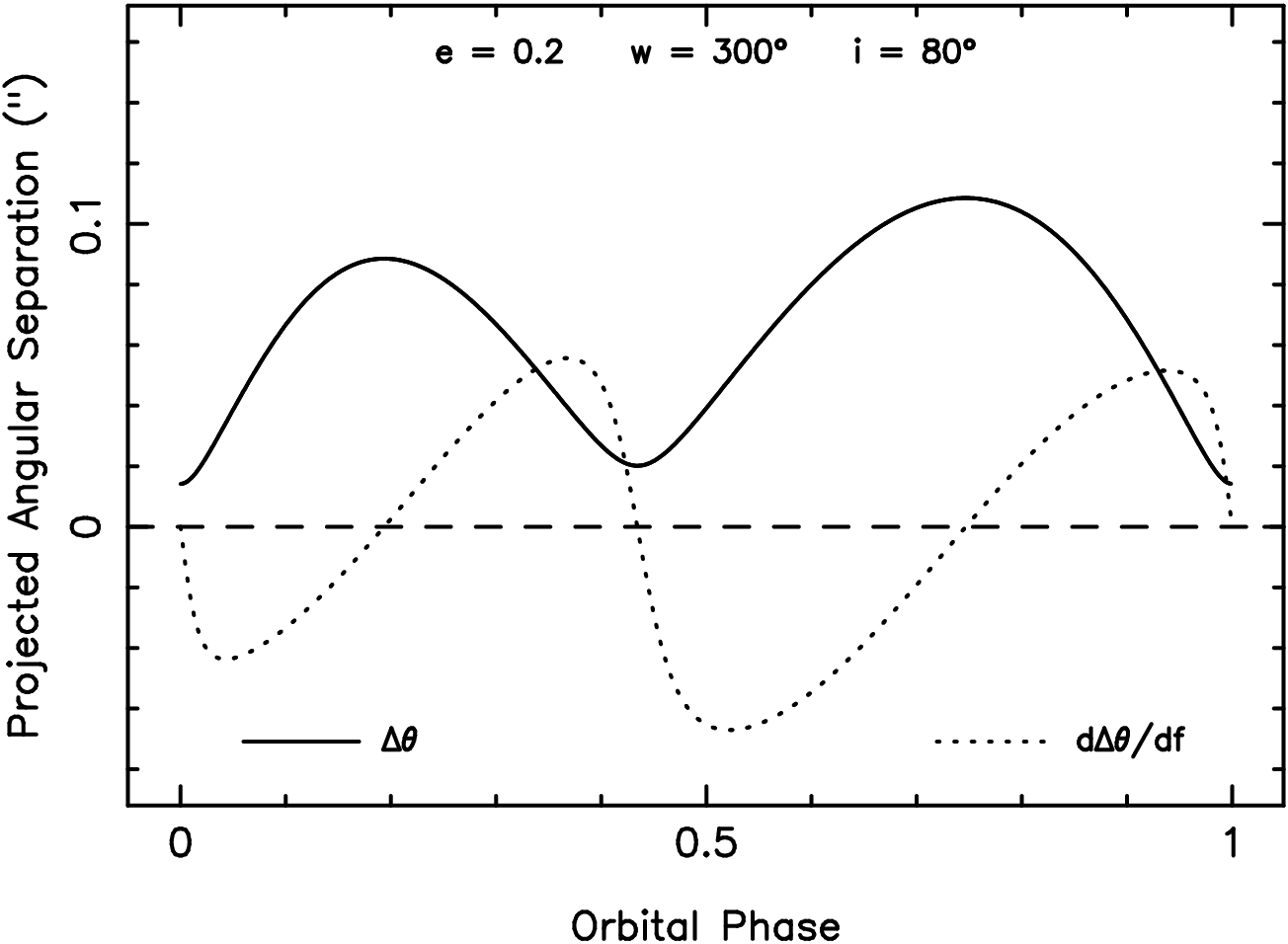} \\
      \includegraphics[angle=270,width=8.5cm]{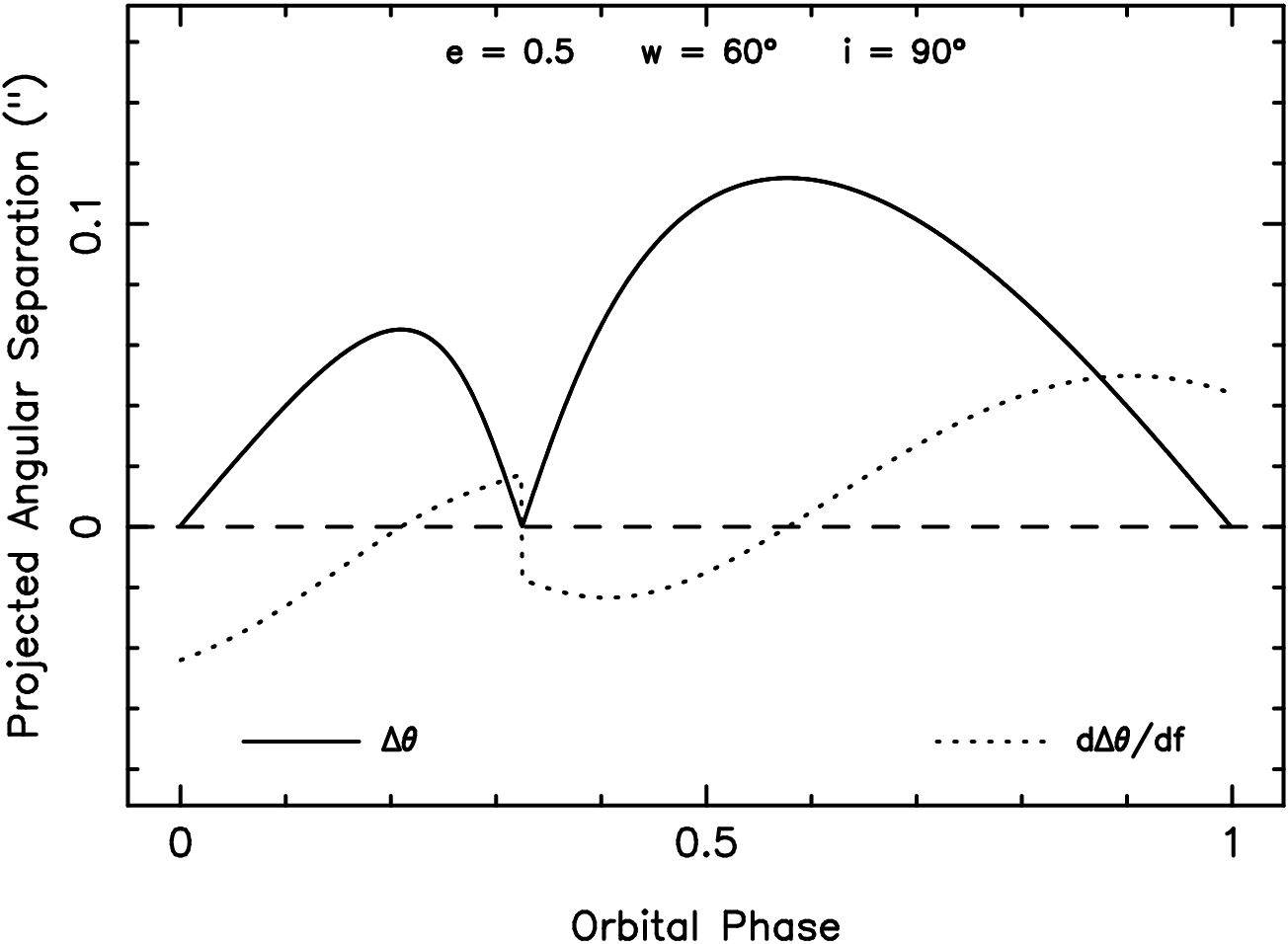} &
      \includegraphics[angle=270,width=8.5cm]{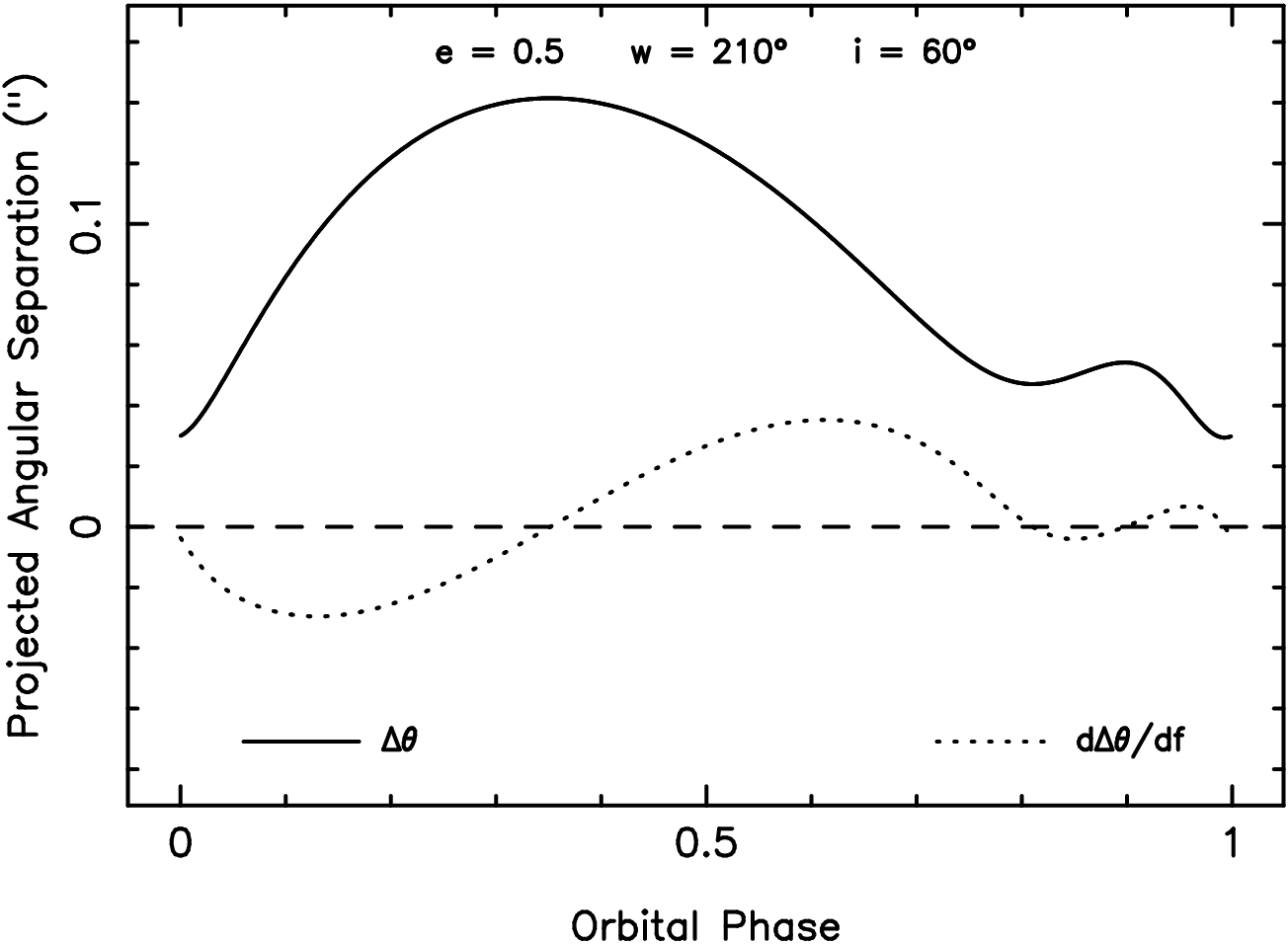}
    \end{tabular}
  \end{center}
  \caption{Projected angular separation (solid line) of a planet in a
    1~AU semi-major axis orbit around a star located 10 parsecs from
    the observer. The four panels shown represent a wide range of
    Keplerian orbital parameters, including eccentricity $e$, argument
    of periastron $\omega$, and orbital inclination $i$. The
    derivative of the angular separation is shown as a dotted line,
    where the intersections with the zero-point (horizontal dashed
    line) indicate the orbital phase locations of the minimum and
    maximum projected angular separation.}
  \label{sepfig}
\end{figure*}

The angular star--planet separation can be sensitive to the Keplerian
orbital element of eccentricity, depending upon the orbital
inclination and the argument of periastron. It is thus critical to
include the full Keplerian orbital solution when calculating the
angular separation.

The star--planet separation, $r$, is generally expressed as
\begin{equation}
  r = \frac{a (1 - e^2)}{1 + e \cos f}
  \label{separation}
\end{equation}
where $a$ is the semi-major axis, $e$ is the eccentricity, and $f$ is
the true anomaly. The angular separation as a function of $f$, as
computed by \citet{kan13}, is given by
\begin{equation}
  \Delta \theta = \frac{r}{d} \left( \cos^2 (\omega + f) + \sin^2
  (\omega + f) \cos^2 i \right)^{\frac{1}{2}}
  \label{angsep}
\end{equation}
where $\omega$ is the argument of periastron, $i$ is the orbital
inclination, and $d$ is the star--observer distance
\citep{kan11}. When expressed in this way, the units of the angular
separation in Equation~\ref{angsep} are in radians.

For the purposes of this work, the main objective is to determine
epochs of maximum angular separation. To achieve that for a Keplerian
orbit, it is necessary to substitute Equation~\ref{separation} into
Equation~\ref{angsep} and differentiate the resulting expression with
respect to $f$. Such differentiation is nontrivial, but does result
in the following analytical expression:
\begin{eqnarray}
  \frac{d}{df} (\Delta \theta) &=& - \frac{a(e-1)^2} {d (e \cos f +
    1)^2} \cdot \nonumber \\
  && \frac{1}{\sqrt{\cos^2 i \sin^2 (\omega+f) + \cos^2 (\omega+f)}}
  \cdot \nonumber \\
  && (e \cos^2 i \sin f \sin^2 (\omega+f) + ((e \cos^2 i - e) \cdot
  \nonumber \\
  && \cos f + \cos^2 i - 1) \cos (\omega+f) \sin (\omega+f) +
  \nonumber \\
  && e \sin f \cos^2 (\omega+f))
  \label{deriv}
\end{eqnarray}
Therefore, the maximum and minimum angular separations occur where
Equation~\ref{deriv} equals zero (stationary points).

Shown in Figure~\ref{sepfig} are four examples of the projected
angular separation (solid line) for a hypothetical planetary system
located 10 parsecs from the observer and with a planetary semi-major
axis of 1~AU. This is demonstrated for eccentricities of 0.2 and 0.5
and for a range of periastron arguments and orbital inclinations. An
orbital phase of zero in the plots corresponds to superior conjunction
where the phase angle is also zero. Also shown in Figure~\ref{sepfig}
as a dotted line is the derivative of the angular separation equation,
corresponding to the rate of change of the angular separation. The
stationary points (where the derivative crosses the zero-point shown
as a horizontal dashed line) indicate the locations of maximum and
minimum angular separations. The maximum angular separations occur at
orbital phases of 0.69, 0.75, 0.58, and 0.35 for the top-left,
top-right, bottom-left, and bottom-right panels respectively. It is
worth noting that the maximum angular separation does not necessarily
occur when the contrast ratio between the star and planet are optimal
for detection, since that also depends on the phase angle and
scattering properties of the atmosphere \citep{kan10,nay17}.

%%%%%%%%%%%%%%%%%%%%%%%%%%%%%%%%%%%%%%%%%%%%%%%%%%%%%%%%%%%%%%%%%%%%

\section{Epochs of Optimal Observation}
\label{epoch}

Here we combine the calculations of the previous two sections and
apply these to the 300 known exoplanets described in
Section~\ref{orbit}. For exoplanets without an eccentricity value, we
fix the orbit to circular ($e = 0.0$). If the argument of periastron
is missing, we fix the periastron to the plane perpendicular to the
sky that aligns with inferior conjunction ($\omega = 90\degr$). For
the majority of the 300 targets considered, the inclination is
unknown, and in those cases, we fix the inclination to an edge-on
orientation ($i = 90\degr$). This inclination was chosen as a
conservative limit since approximately edge-on orbits are the most
difficult for direct imaging detection.

Table~\ref{epochtab} shows the top 50 known exoplanet targets ranked
by their maximum angular separation, $\Delta \theta_\mathrm{max}$,
shown in units of milliarcsecs (mas). Also included are the predicted
orbital phase past superior conjunction (phase angle of zero) where
the maximum angular separation will occur, $\phi_\mathrm{max}$, and
the uncertainty (in orbital phase units) of when that will occur,
$\sigma_\phi$. All of the planets represented in Table~\ref{epochtab}
are assumed to have edge-on orbits with the exception of eps~Eri~b,
which has a measured orbital inclination of $i = 30.1\degr$
\citep{ben06}. Therefore, the planetary masses, $M_p$, are minimum
masses (except for eps~Eri~b) in units of Jupiter masses, $M_J$. Note
that the values of $\Delta \theta_\mathrm{max}$ and
$\phi_\mathrm{max}$ do not change with time (unless the orbital
solution is updated), but the uncertainty in phase, $\sigma_\phi$,
where $\Delta \theta_\mathrm{max}$ occurs does increase with time and
is calculated for 2025, as described in Section~\ref{orbit}. This
means that the $\sigma_\phi$ values apply to the next maximum angular
separation event that occurs past the 2025 date.

\begin{deluxetable*}{lcccccccccc}
  \tablewidth{0pc}
  \tablecaption{\label{epochtab} Maximum angular separations.}
  \tablehead{
    \colhead{Planet} &
    \colhead{$P$} &
    \colhead{$a$} &
    \colhead{$e$} &
    \colhead{$\omega$} &
    \colhead{$i$} &
    \colhead{$M_p$} &
    \colhead{$d$} &
    \colhead{$\Delta \theta_\mathrm{max}$} &
    \colhead{$\phi_\mathrm{max}$} &
    \colhead{$\sigma_\phi$} \\
    \colhead{} &
    \colhead{(days)} &
    \colhead{(AU)} &
    \colhead{} &
    \colhead{($\degr$)} &
    \colhead{($\degr$)} &
    \colhead{($M_J$)} &
    \colhead{(pcs)} &
    \colhead{(mas)} &
    \colhead{} &
    \colhead{}
  }
  \startdata
eps Eri b            &  2502.00 &  3.39 & 0.70 &  47.0 & 30.1 &  1.55 &    3.2 & 1679.6 &   0.609 &   0.013 \\
47 UMa d             & 14002.00 & 11.60 & 0.16 & 110.0 & 90.0 &  1.64 &   14.1 &  859.6 &   0.293 &   0.630 \\
HD 217107 c          &  4270.00 &  5.32 & 0.52 & 198.6 & 90.0 &  2.60 &   19.7 &  398.3 &   0.367 &   0.125 \\
GJ 676 A c           &  7337.00 &  6.60 & 0.00 &  90.0 & 90.0 &  6.80 &   16.9 &  390.5 &   0.249 &   0.026 \\
HD 160691 c          &  4205.80 &  5.24 & 0.10 &  57.6 & 90.0 &  1.81 &   15.3 &  359.5 &   0.719 &   0.455 \\
HD 150706 b          &  5894.00 &  6.70 & 0.38 & 132.0 & 90.0 &  2.71 &   27.2 &  298.6 &   0.387 &   0.762 \\
HD 134987 c          &  5000.00 &  5.80 & 0.12 & 195.0 & 90.0 &  0.82 &   22.2 &  291.4 &   0.281 &   0.113 \\
HD 142 c             &  6005.00 &  6.80 & 0.21 & 250.0 & 90.0 &  5.30 &   25.6 &  279.0 &   0.239 &   0.100 \\
47 UMa c             &  2391.00 &  3.60 & 0.10 & 295.0 & 90.0 &  0.54 &   14.1 &  265.3 &   0.751 &   0.439 \\
HD 219077 b          &  5501.00 &  6.22 & 0.77 &  57.6 & 90.0 & 10.39 &   29.2 &  249.7 &   0.452 &   0.042 \\
GJ 328 b             &  4100.00 &  4.50 & 0.37 & 290.0 & 90.0 &  2.30 &   20.0 &  239.1 &   0.768 &   0.135 \\
GJ 179 b             &  2288.00 &  2.41 & 0.21 & 153.0 & 90.0 &  0.82 &   12.1 &  235.1 &   0.326 &   0.108 \\
HD 166724 b          &  5144.00 &  5.42 & 0.73 & 202.3 & 90.0 &  3.53 &   43.0 &  206.7 &   0.400 &   0.273 \\
ups And d            &  1276.46 &  2.51 & 0.30 & 258.8 & 90.0 &  4.13 &   13.5 &  189.2 &   0.219 &   0.004 \\
HD 196067 b          &  3638.00 &  5.02 & 0.66 & 148.2 & 90.0 &  6.90 &   43.6 &  172.7 &   0.499 &   0.177 \\
HD 113538 c          &  1818.00 &  2.44 & 0.20 & 280.0 & 90.0 &  0.93 &   15.9 &  155.8 &   0.771 &   0.045 \\
HAT-P-11 c           &  3407.00 &  4.13 & 0.60 & 143.7 & 90.0 &  1.60 &   37.8 &  155.0 &   0.480 &   0.097 \\
47 UMa b             &  1078.00 &  2.10 & 0.03 & 334.0 & 90.0 &  2.53 &   14.1 &  153.4 &   0.742 &   0.080 \\
nu Oph c             &  3186.00 &  6.10 & 0.17 &   4.6 & 90.0 & 27.00 &   46.8 &  151.6 &   0.695 &   0.022 \\
HD 106515 A b        &  3630.00 &  4.59 & 0.57 & 123.8 & 90.0 &  9.61 &   36.4 &  151.2 &   0.459 &   0.010 \\
gam Cep b            &   903.30 &  2.05 & 0.05 &  94.6 & 90.0 &  1.85 &   13.8 &  149.1 &   0.259 &   0.110 \\
HD 141399 e          &  5000.00 &  5.00 & 0.26 &  90.0 & 90.0 &  0.66 &   36.2 &  133.4 &   0.708 &   0.251 \\
HD 10647 b           &   989.20 &  2.02 & 0.15 & 212.0 & 90.0 &  0.94 &   17.4 &  130.5 &   0.275 &   0.158 \\
HD 220773 b          &  3724.70 &  4.94 & 0.51 & 226.0 & 90.0 &  1.45 &   49.0 &  129.5 &   0.293 &   0.253 \\
HD 133131 B b        &  5769.00 &  6.15 & 0.61 & 103.0 & 90.0 &  2.50 &   47.0 &  123.2 &   0.413 &   0.394 \\
HD 98649 b           &  4951.00 &  5.60 & 0.85 & 248.0 & 90.0 &  6.80 &   42.8 &  122.3 &   0.244 &   0.232 \\
HD 219828 c          &  4791.00 &  5.96 & 0.81 & 145.8 & 90.0 & 15.10 &   77.9 &  119.4 &   0.548 &   0.021 \\
HD 8673 b            &  1634.00 &  3.02 & 0.72 & 323.4 & 90.0 & 14.20 &   38.3 &  117.0 &   0.646 &   0.045 \\
HD 38529 c           &  2140.20 &  3.71 & 0.34 &  17.8 & 90.0 & 13.38 &   42.4 &  115.5 &   0.627 &   0.013 \\
HD 187123 c          &  3810.00 &  4.89 & 0.25 & 243.0 & 90.0 &  1.99 &   47.9 &  111.1 &   0.247 &   0.205 \\
HD 160691 b          &   643.25 &  1.50 & 0.13 &  22.0 & 90.0 &  1.08 &   15.3 &  109.5 &   0.704 &   0.038 \\
HD 29021 b           &  1362.30 &  2.28 & 0.46 & 179.5 & 90.0 &  2.40 &   30.6 &  108.7 &   0.391 &   0.016 \\
HD 147513 b          &   528.40 &  1.32 & 0.26 & 282.0 & 90.0 &  1.21 &   12.9 &  104.7 &   0.774 &   0.253 \\
HD 169830 c          &  2102.00 &  3.60 & 0.33 & 252.0 & 90.0 &  4.04 &   36.3 &  104.2 &   0.229 &   0.499 \\
HD 4203 c            &  6700.00 &  6.95 & 0.24 & 224.0 & 90.0 &  2.17 &   77.8 &  103.5 &   0.274 &   1.902 \\
HD 183263 c          &  3070.00 &  4.35 & 0.24 & 345.0 & 90.0 &  3.57 &   52.8 &  101.2 &   0.688 &   0.132 \\
HD 181433 d          &  2172.00 &  3.00 & 0.48 &  90.0 & 90.0 &  0.54 &   26.1 &  100.6 &   0.325 &   0.375 \\
7 CMa b              &   796.00 &  1.93 & 0.22 &  77.0 & 90.0 &  2.46 &   19.8 &   99.8 &   0.698 &   0.120 \\
HD 32963 b           &  2372.00 &  3.41 & 0.07 & 107.0 & 90.0 &  0.70 &   35.2 &   98.5 &   0.267 &   0.183 \\
HD 216437 b          &  1256.00 &  2.32 & 0.29 &  63.0 & 90.0 &  1.82 &   26.5 &   96.0 &   0.660 &   0.325 \\
HD 10180 h           &  2205.00 &  3.38 & 0.09 & 142.0 & 90.0 &  0.21 &   39.0 &   93.0 &   0.283 &   0.336 \\
HD 11964 b           &  1945.00 &  3.16 & 0.04 &  90.0 & 90.0 &  0.62 &   34.0 &   92.9 &   0.256 &   0.240 \\
HD 133131 A c        &  3568.00 &  4.49 & 0.49 & 100.0 & 90.0 &  0.42 &   47.0 &   91.8 &   0.365 &   0.717 \\
HD 37605 c           &  2720.00 &  3.81 & 0.01 & 221.0 & 90.0 &  3.37 &   42.9 &   89.8 &   0.251 &   0.259 \\
HD 4732 c            &  2732.00 &  4.60 & 0.23 & 118.0 & 90.0 &  2.37 &   56.5 &   88.5 &   0.321 &   0.087 \\
HD 79498 b           &  1966.10 &  3.13 & 0.59 & 221.0 & 90.0 &  1.34 &   49.0 &   87.3 &   0.317 &   0.099 \\
HAT-P-17 c           &  5584.00 &  5.60 & 0.39 & 181.5 & 90.0 &  3.40 &   90.0 &   86.5 &   0.369 &   0.919 \\
BD-11 4672 b         &  1667.00 &  2.28 & 0.05 & 231.0 & 90.0 &  0.53 &   27.3 &   86.1 &   0.253 &   0.080 \\
HD 181433 c          &   962.00 &  1.76 & 0.28 &  21.4 & 90.0 &  0.64 &   26.1 &   84.5 &   0.647 &   0.128 \\
GJ 317 b             &   692.00 &  1.15 & 0.11 & 342.0 & 45.0 &  2.50 &   15.1 &   84.1 &   0.717 &   0.088 \\
  \enddata
\end{deluxetable*}

\begin{figure*}
  \begin{center}
    \begin{tabular}{cc}
      \includegraphics[angle=270,width=7.2cm]{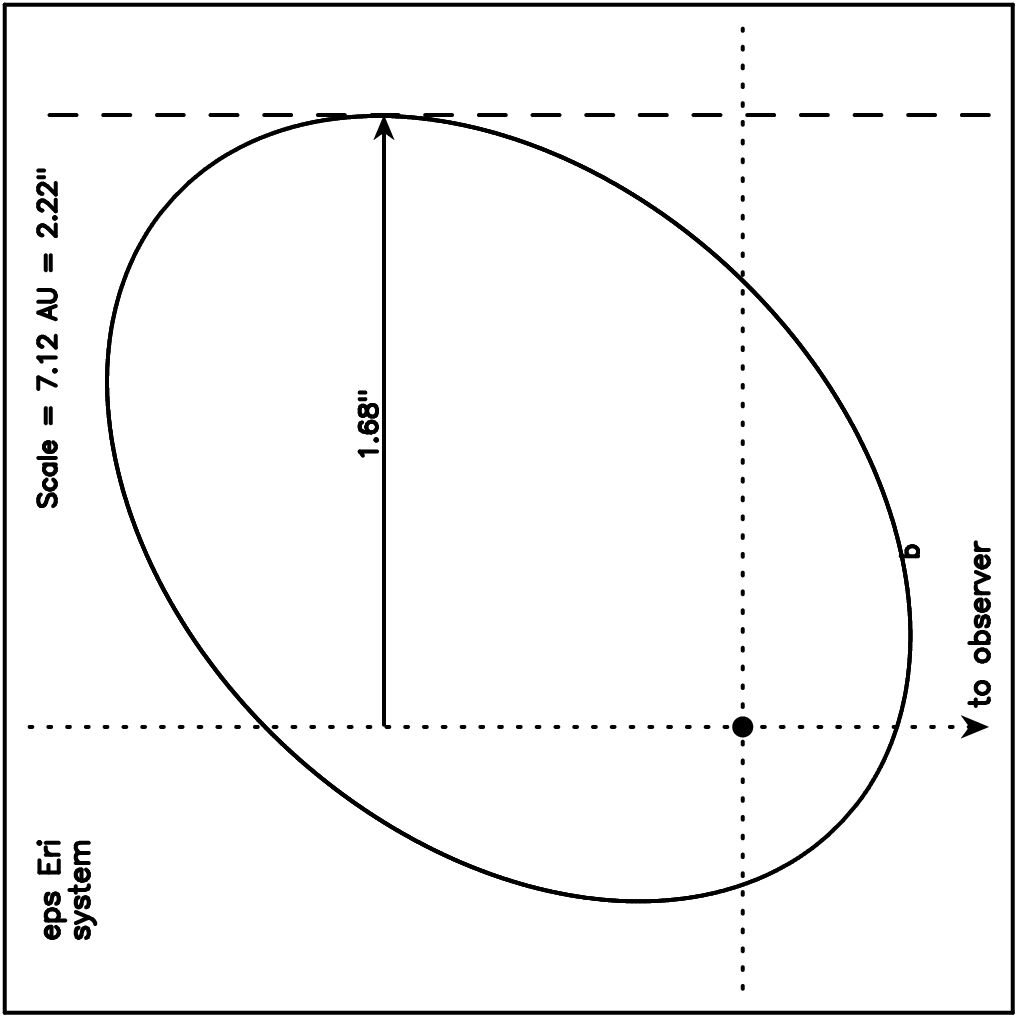} &
      \includegraphics[angle=270,width=9.8cm]{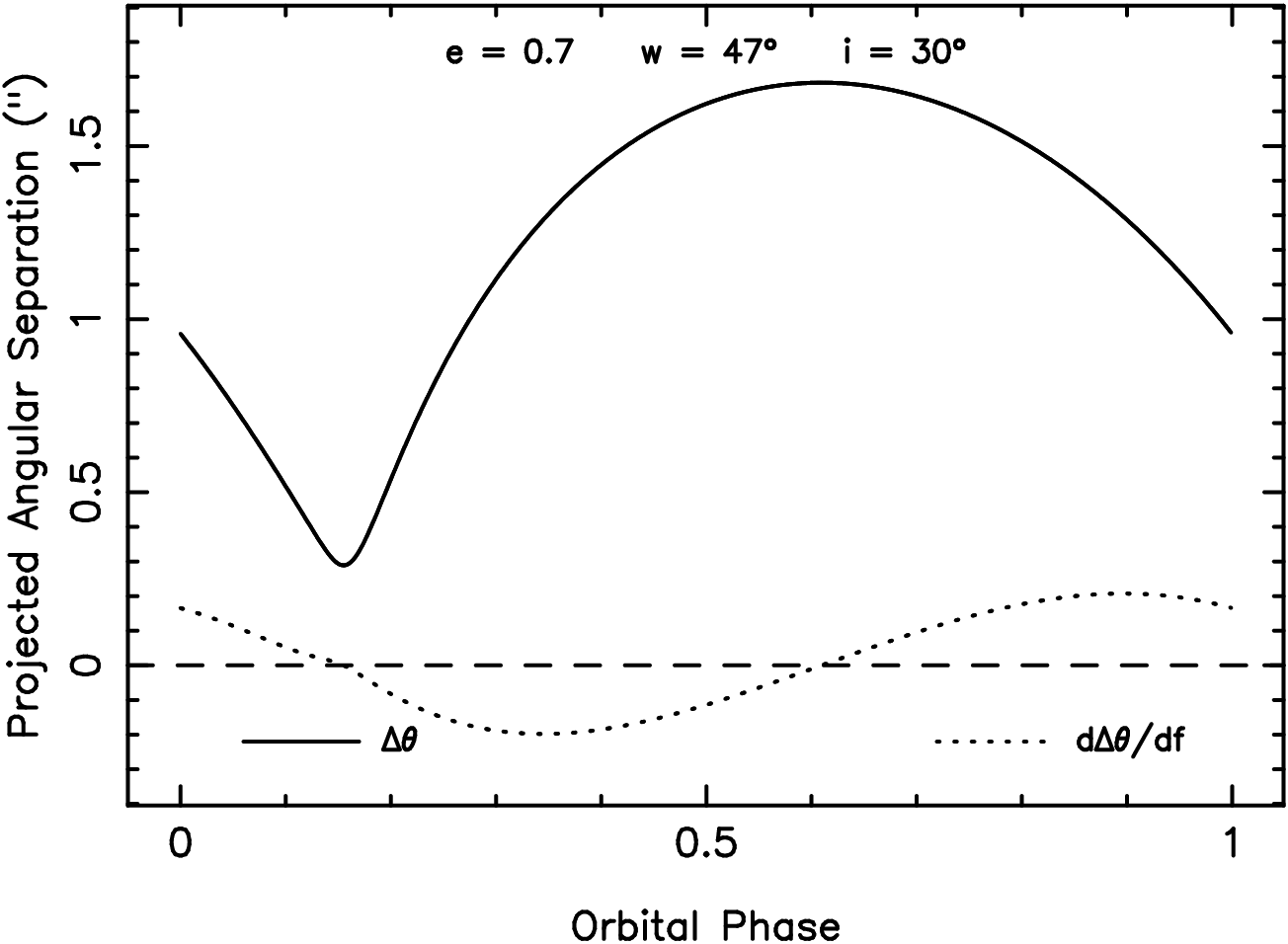}
    \end{tabular}
  \end{center}
  \caption{Plots showing the orbit and angular separation for eps Eri
    b. The left panel represents a top-down view of the eps Eri b
    orbit where the orientation with respect to the observer is shown
    with the dotted lines. The size of the plot one side is equivalent
    7.12~AU which, as the distance of the system ($d = 3.21$~pcs) is
    equivalent to an angular size of 2.22\arcsec. The vertical dashed
    line indicates the location of maximum angular separation of the
    planet from the host star as seen from the observer, corresponding
    to the 1.68\arcsec value shown in Table~\ref{epochtab}. The right
    panel shows the variation in maximum angular separation (solid
    line) and rate of angular separation change (dotted line) for eps
    Eri b.}
  \label{epserifig}
\end{figure*}

The highest ranked case of eps~Eri~b within Table~\ref{epochtab} is
represented in the panels of Figure~\ref{epserifig}. The left panel
shows a top-down view of the orbit, which is highly eccentric ($e =
0.7$) and has a predicted maximum angular separation of $\Delta
\theta_\mathrm{max} = 1.68\arcsec$. The right panel displays the
angular separation as well as the rate of angular separation change
(derivative of angular separation, see Equation~\ref{deriv}), as
described in Figure~\ref{sepfig}. The combination of the large
predicted maximum angular separation and the relatively small
uncertainty on the orbital ephemeris (0.013 phase units) make this an
ideal target for follow-up observations from an orbit
perspective. More recent work by \citet{maw18} suggests that eps~Eri~b
has a substantially more circular orbit ($e = 0.07$), which would
reduce the predicted maximum angular separation to $\Delta
\theta_\mathrm{max} = 1.14\arcsec$ but maintain the planet's
top-ranked position in Table~\ref{epochtab}. The well-defined orbit,
including orbital inclination, is a result of a simultaneous fit of RV
and astrometric data through the use of Hubble Space Telescope (HST)
observations \citep{ben06}. Note that eps~Eri is an active star that
will present other observational challenges for direct detection of
the known exoplanet \citep{met13,jef14}.

By contrast, cases such as the 47~UMa system have large predicted
maximum angular separations but relatively large uncertainties
concerning when that separation will occur (0.439 and 0.630 phase
units for the c and d planets, respectively). Such systems will
benefit enormously from further RV observations at specific epochs
that will provide vast improvements to the orbital solution
\citep{kan09}. Provided that the uncertainty in orbital phase can be
constrained to cover a range of orbital locations that lie outside the
inner working angle of an instrumental design, then the targets will
be viable for observations.

%%%%%%%%%%%%%%%%%%%%%%%%%%%%%%%%%%%%%%%%%%%%%%%%%%%%%%%%%%%%%%%%%%%%

\section{Conclusions}
\label{conclusions}

A key component of designing imaging missions is the selection of
optimal targets for observation. These are naturally drawn from the
known RV exoplanets since these provide test cases for technology
demonstrations and contain the necessary long-period demographic
required by direct imaging experiments. The Keplerian nature of the RV
orbits can lead to enhanced angular separations, though the timing of
such separations is often poorly constrained. The methodology provided
here allows the direct calculation of maximum angular separation via
the stationary points of the angular separation equation. As stated at
the end of Section~\ref{ang}, the epochs of maximum angular separation
do not necessarily correspond with the epochs of expected maximum
planet brightness. The contrast ratio of exoplanets depends upon
numerous factors such as the wavelength of observation and also the
type (terrestrial, gas giant), age, atmospheric properties, and albedo
of the planet \citep{fen18}. The focus of this work is to allow the
observation of planets that would otherwise be inside of the inner
working angle of the imaging experimental design \citep{tur12}.

The challenge of improving the RV targets to ensure that they will
minimize telescope resources is one that must be met before a
systematic imaging survey can commence. The RV time required to refine
the orbits of long-period planets can be moderate, provided that one
utilizes the same facility that was used to acquire the discovery data
\citep{kan09}. Precise observing strategies depend on the properties
of the individual targets and need to be customized on a case-by-case
basis \citep{kan07,bot13}. Refining the orbits of the planets
discussed in this paper will help enormously toward increasing the
detection yield of missions such as WFIRST, HabEx, and LUVOIR and will
also aid in planning follow-up observations with {\it James Webb Space
  Telescope} for detecting phase variations of known planets. For
missions launching in the mid 2020s, it is paramount that the process
of orbital refinement commences with sufficient lead time to avoid
compromising the target list.

%%%%%%%%%%%%%%%%%%%%%%%%%%%%%%%%%%%%%%%%%%%%%%%%%%%%%%%%%%%%%%%%%%%%

\section*{Acknowledgements}

This research has made use of the NASA Exoplanet Archive, which is
operated by the California Institute of Technology, under contract
with the National Aeronautics and Space Administration under the
Exoplanet Exploration Program. The results reported herein benefited
from collaborations and/or information exchange within NASA's Nexus
for Exoplanet System Science (NExSS) research coordination network
sponsored by NASA's Science Mission Directorate. This work was funded
by the WFIRST CGI Science Investigation Team contract \#NNG16P27C (PI:
Margaret Turnbull).

%%%%%%%%%%%%%%%%%%%%%%%%%%%%%%%%%%%%%%%%%%%%%%%%%%%%%%%%%%%%%%%%%%%%

\end{document}